\documentclass{article}
\usepackage{graphicx}
\usepackage{float}
\usepackage{tabularx}
\usepackage{multirow}
\usepackage{booktabs}
\usepackage[a4paper, total={6in, 9in}]{geometry}

\title{School Bullying Results in Poor Psychological Conditions: Evidence from a Survey of 95,545 Subjects}

\author{Na Zhao$^{1,2}$ \quad Shenglong Yang$^{1}$ \quad Qiangjian Zhang$^{1}$ \quad  Jian Wang$^{3}$\quad  Wei Xie$^{4}$  \\ \quad Youguo Tan$^{5}$ \quad Tao Zhou$^{2,6}$\footnotemark[1]
\\
$^{1}$Yunnan University \quad $^{2}$SeekingTao Tech. Inc. \quad  $^{3}$Kunming University of Science and Technology \quad  
\\ $^{4}$Chengdu Happy Xiaoqing Intelligent Technology Co. LTD \quad\\
$^{5}$Zigong Fifth People's Hospital \quad
$^{6}$University of Electronic Science and Technology of China \quad \\
$^{*}$Correspondence: zhutou@ustc.edu
}

\begin{document}
\date{}
\maketitle

\abstract{To investigate whether bullying and psychological conditions are correlated, this study analyzed a survey of primary and secondary school students from Zigong City, Sichuan Province. A total of 95,545 students completed a personal information questionnaire, the Multidimensional Peer-Victimization Scale (MPVS), and eight other scales pertaining to various psychological problems. The data showed that 68,315 (71.5\%) participants experienced school bullying at varying degrees, indicating the prevalence of bullying among adolescents. The chi-square tests revealed a strong correlation between school bullying and psychological conditions. This correlation was further explored through multivariate logistic regression, showing that students who experienced mild bullying had a 3.10 times higher probability of emotional and behavioral problems, 4.06 times higher probability of experiencing prodromal symptoms of mental illness, 4.72 times higher probability of anxiety, 3.28 times higher probability of developing post-traumatic stress disorder (PTSD) , 4.07 times higher probability of poor sleep quality, 3.13 times higher probability of internet addiction, 2.18 times higher probability of poor mental health, and 3.64 times higher probability of depression than students who did not experience bullying. The corresponding probabilities for students who experienced severe bullying were 11.35, 17.35, 18.52, 12.59, 11.67, 12.03, 4.64, and 5.34 times higher, respectively. In conclusion, school bullying and psychological conditions are significantly correlated among primary and secondary school students, and the more severe the bullying, the higher the probability to suffer from psychological problems.

}

\section{Introduction}The teenage years are a critical period for physical growth and development, mental maturity, personality formation, and the attainment of scientific and cultural knowledge. Psychological problems, such as depression, can lead to several negative outcomes in teenagers, including poor academic performance, alcohol abuse, and suicide \cite{ref1}. The reasons for poor psychological conditions are complex, involving personal, family, and school factors. Schools, as large gathering places for primary and secondary school students, allow students to frequently interact with each other. These interactions, including school bullying, can significantly impact students’ psychological well-being \cite{2}.
\par
The increasing number of cases of school bullying in the 21st century has drawn widespread social attention to related studies \cite{3}. Early studies showed that psychological conditions are strongly related to school bullying, which is defined as continued attacks by teachers or students on certain students within the school campuses and surrounding areas \cite{4}. Traditional bullying can be divided into four categories: physical victimization, verbal victimization, social manipulation, and attacks on property \cite{5}. School bullying is prevalent among teenagers globally: the proportion of teenagers affected by school bullying across different countries varies between 4.8\% and 45.2\% \cite{6}. 
\par
Researchers have conducted in-depth studies on school bullying. For example, Cosma \textit{et al}. \cite{7} compared the cross-national trends of school bullying to examine the differences between cyberbullying and traditional bullying. Chudal \textit{et al}. \cite{8} investigated the prevalence of bullying and its various types among 21,688 13–15-year-old adolescents in developing countries. Pengpid \textit{et al}. \cite{9} studied the relationship between school bullying and psychological problems among adolescents in Southeast Asian countries. Pörhölä \textit{et al}. \cite{10} surveyed 8,497 students from several countries to examine the differences between various types of school bullying. Chen \textit{et al}. \cite{11} analyzed 4,051 bullied adolescents, revealing the relationship between school bullying and post-traumatic stress disorder (PTSD). Meanwhile, Zhou \textit{et al}. \cite{12} systematically studied the correlation between school bullying and poor sleep quality. However, the existing body of literature on the relationship between school bullying and psychological problems has shortcomings, including small sample sizes, lack of attention paid to adolescents in developing countries, lack of comparative studies on various types of bullying, and lack of analysis on multiple dimensions of psychological problems.
\par
The present study aimed to fill this gap in the literature by surveying a large sample of students from developing countries. Furthermore, to offer detailed findings, this study sought to cover the four types of school bullying and distinguish between different degrees and different types of school bullying.
\section{Methods}
\subsection{Participants}
This study involved a questionnaire survey of 95,545 primary and secondary school students. The sample included 27,128 (29\%) primary school students from 71 primary schools, 43,124 (45\%) junior high school students from 73 junior high schools, and 25,203 (26\%) senior high school students from 14 senior high schools in Zigong City, Sichuan Province, China. The participants were aged 6–22 years (average age: 13.47 years), with primary school students mainly 6–12 years, junior high school students mainly 13–15 years, and senior high school students mainly 16–18 years.

\subsection{Ethical Statement}
All participants were informed of the purpose of the study, and participation was voluntary. Each participant and her/his parents signed written consent forms. To protect the participants’ privacy, the survey data were anonymized. 

\subsection{Questionnaire Survey Data}
\textit{Basic information}. Using the Basic Information Survey Questionnaire, the study collected data on participants’ age, school, living environment, family situation, and family environment, to analyze the reasons for their psychological problems from personal and family perspectives.
\par
\textit{School bullying}. The Multidimensional Peer-Victimization Scale (MPVS) \cite{5} was used to investigate the prevalence of school bullying, including physical and verbal victimization, property damage, and insults from peers during the participants' growth process. The types of victimization included hitting, insulting, teasing, slandering, exclusion, spreading rumors, and so on. The participants were required to truthfully fill out the MPVS to reflect their experiences of school bullying.
\par
\textit{Psychological problems}. We assessed participants’ overall psychological well-being by measuring their psychological health status across eight dimensions. These dimensions corresponded to validated scales, including the Strengths and Difficulties Questionnaire - Short Version (SDQ-S) \cite{13} for emotional and behavioral problems, the Prodromal Questionnaire - 16 Items (PQ-16) \cite{14} for psychotic risk, the Generalized Anxiety Disorder 7-Item Scale (GAD-7) \cite{15} for anxiety, the Children’s Revised Impact of Event Scale – 13 Items (CRIES-13) \cite{16} for stress response, the Pittsburgh Sleep Quality Index (PSQI) \cite{17, 18} for sleep quality, the nine-item Internet Gaming Disorder Scale - Short Form (IGDS9-SF) \cite{19, 20} for internet addiction, the Patient Health Questionnaire - 9 item (PHQ-9) \cite{21} for depression, and the Warwick-Edinburgh Mental Well-being Scale (WEMWBS) \cite{15} for mental well-being. After participants completed the aforementioned scales, we evaluated participants’ psychological conditions based on the scale scores.

\subsection{Severity of Psychological Problems}
The students were classified according to the diagnostic criteria in Table \ref{table_class}, based on the results of the psychological health survey. All subsequent investigations in this study were conducted based on those criteria.

\begin{table}[H] 
\renewcommand{\arraystretch}{1.4}
\caption{Classification of the severity of psychological problems, where * represents that the respondent is considered to be in a state of illness. The regression analysis in Section 3.6 below is based on this dichotomy of "ill - not ill".\label{table_class}}
\newcolumntype{C}{>{\centering\arraybackslash}X}
\begin{tabularx}{\textwidth}{p{1.5cm}p{1.5cm}p{6cm}p{5cm}}
\toprule
Scale  & Total Score & Symptom  & Criterion \\
\midrule
\multirow{3}{*}{SDQ-S} & \multirow{3}{*}{40} & 0-Healthy & 0 $\leq$ score $\leq$ 15  \\ 
& & 1-Moderate* & 16 $\leq$ score $\leq$ 19  \\
& & 2-Heavy* & 20 $\leq$ score $\leq$ 40 \\
\hline
\multirow{2}{*}{PQ-16} & \multirow{2}{*}{48} & 0-No risk of psychosis & score $\geq$ 10 \\ 
& & 1-Risk for psychosis* & score \textless  10  \\
\hline
\multirow{4}{*}{GAD-7} & \multirow{4}{*}{21} & 0-No anxiety &  0 $\leq$ score \textless 5\\
 &  & 1-Mild anxiety*        & 5 $\leq$ score \textless 10                    \\
 &    & 2-Moderate anxiety*       & 10 $\leq$ score \textless 15                   \\
 &   & 3-Severe anxiety*        & 15 $\leq$ score \\ 
 \hline
\multirow{3}{*}{CRIES-13} & \multirow{3}{*}{65} & 0-Healthy     & score $\leq$ 17 \\
&  & 1-Moderate*   & 18 $\leq$ score $\leq$  31                 \\
 &  & 2-Heavy*    & 32 $\leq$ score                  \\ 
 \hline
\multirow{4}{*}{PSQI}  & \multirow{4}{*}{21}   & 0-Very good sleep quality &score = 0  \\
 &  & 1-Fairly good sleep quality  & 1 $\leq$ score \textless 11                 \\
 &  & 2-Poor sleep quality* & 11 $\leq$ score \textless 16                 \\
 &  & 3-Very poor sleep quality* & 16 $\leq$ score            \\ 
 \hline
\multirow{2}{*}{IGDS9-SF} & \multirow{2}{*}{45} & 0-No internet addiction     & questions that scores 5 points \textless 5            \\
 &  & 1-Internet addiction*    & question that scores 5 points $\geq$ 5                   \\ 
\hline
\multirow{5}{*}{PHQ-9} & \multirow{5}{*}{27} & 0-Healthy  & 0 $\leq$ score \textless 5  \\ 
&   & 1-Mild depression*            & 5 $\leq$ score \textless 10                    \\
&   & 2-Moderate depression*        & 10 $\leq$ score \textless 15                   \\
&   & 3-Moderately severe depression*       & 15 $\leq$ score \textless 20                   \\
&   & 4-Severe depression*    & 20 $\leq$ score   \\
\hline
\multirow{4}{*}{WEMWBS}  & \multirow{4}{*}{70} & 0-Very low level of mental health*  & 14 $\leq$score \textless 32 \\
 &   & 1-Low level of mental health* & 32 $\leq$ score \textless 40                 \\
 &   & 2-Moderate level of mental health & 40 $\leq$ score \textless 59                 \\
 &   & 3-Very high level of mental health & 59 $\leq$ score $\leq$ 70                \\ 
\bottomrule
\end{tabularx}
\end{table}

\subsection{Statistical Methods}
Based on the classification of psychological problems, as described above, we performed chi-square tests to validate the significance of the correlations between different variables, including the correlations between school bullying and the presence of psychological problems and between the degree of bullying and the presence of psychological problems. After confirming the correlations between school bullying and psychological problems, we further performed logistic regression to quantitatively analyze those correlations. For all the results from the statistical analyses, a p-value less than 0.05 was considered to indicate the statistical significance.

\section{Results}
\subsection{Overview of School Bullying}
\begin{figure}[H]
\centering
\includegraphics[width=10.5 cm]{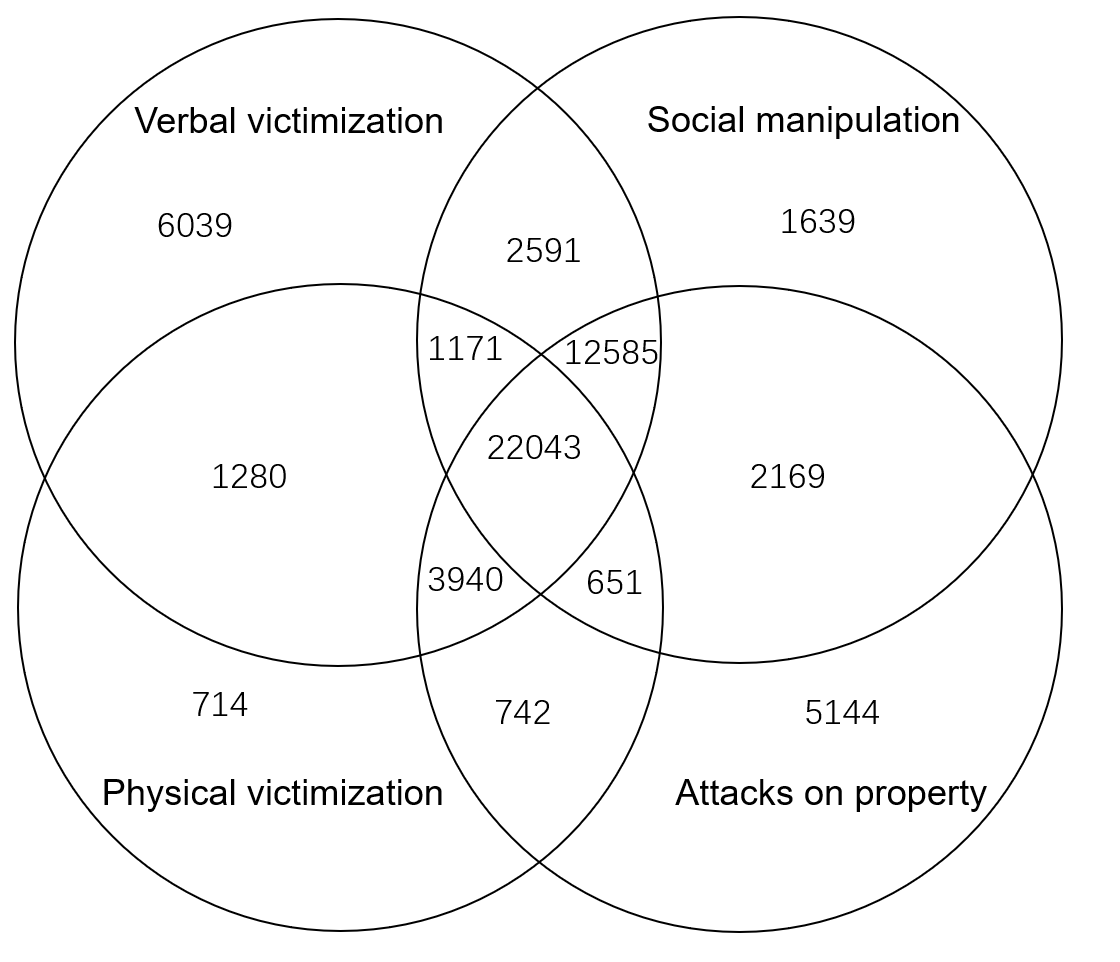}
\caption{A Venn diagram of types of school bullying. The Venn diagram illustrates the number of students who have experienced different types of school bullying. Each block is marked by the number of students in the corresponding set. For example, 1171 represents the number of students who have experienced verbal victimization, physical victimization and social manipulation, but not attacks on property. Two data points are not shown in the diagram: 7455 students who experienced both verbal victimization and attacks on property and 179 students who experienced both physical victimization and social manipulation. \label{fig_vens}}
\end{figure}

Of the 95,545 primary and secondary school students surveyed, 68,342 (71.6\%) reported that they experienced school bullying of varying types and degrees. Based on the MPVS scores, we classified students who experienced school bullying into two categories by the degree of victimization: mild victimization (MPVS score less than or equal to 16) and severe victimization (MPVS score larger than 16). Among those who experienced school bullying, 56,372 and 11,970 students experienced mild and severe bullying, accounting for 59\% and 12.5\% of the total sample, respectively (see Table \ref{table_all}). Table \ref{table_all} reveals that there are overlaps among different types of bullying, namely some students may have experienced multiple types of bullying, as indicated in Figure \ref{fig_vens}.

\begin{table}[H]
\renewcommand{\arraystretch}{1.5}
\caption{The overall statistics of school bullying in this survey.\label{table_all}}
\newcolumntype{C}{>{\centering\arraybackslash}X}
\begin{tabularx}{\textwidth}{p{4cm}p{4cm}p{2cm}p{1.5cm}}
\toprule
  & Category  & Samples   & Percentage\\
\midrule
\multirow{2}{*}{Whether being bullying}   & Being bullied  & 68342 & 71.6\%  \\
                          & Not being bullied & 27113 & 28.4\%  \\ \hline
\multirow{4}{*}{Types of school bullying}   & Physical victimization   & 30720 & 32.2\%  \\
                          & Verbal victimization   & 57104 & 59.8\%  \\
                          & Social manipulation   & 43028 & 45.10\% \\
                          & Attacks on property   & 54729 & 57.30\% \\ \hline
\multirow{2}{*}{Severity of school bullying} & Mild bullying & 56372 & 59.00\% \\
                          & evere bullying & 11970 & 12.50\% \\
\bottomrule
\end{tabularx}
\end{table}

\unskip

\subsection{Overall Impact of Being Bullied}
Table \ref{table1} lists the eight dimensions of psychological problems. Among all participants, the prevalence of severe emotional and behavioral problems (SDQ-S) was 11.4\%; 16.5\% were found to be at risk of mental illness (PQ-16); 12.9\% reported moderate to severe anxiety (GAD-7); 22.6\% showed relatively strong stress responses, with 4.9\% at risk of developing PTSD (CRIES-13); 6.1\% of students had poor or very poor sleep quality (PSQI); 7.7\% experienced internet addiction (IGDS9-SF); 16.2\% had moderate to severe depression (PHQ-9); and 28.1\% had low or very low levels of mental health (WEMWBS). These findings indicate that the psychological conditions of Chinese primary and secondary school students are poor as a whole. The results of the chi-square tests showed significant correlations between experiencing bullying and all eight psychological problems, that is, students who were bullied had significantly poorer psychological conditions than those who were not.

\begin{table}[H]
\renewcommand{\arraystretch}{1.5}
\caption{The impact of school bullying on psychological conditions (*: \textless0.1, **: \textless0.05, ***: \textless0.01, the same below). \label{table1}}
\newcolumntype{C}{>{\centering\arraybackslash}X}
\begin{tabularx}{\textwidth}{p{2cm}p{1.5cm}p{2cm}p{2cm}p{2cm}p{2cm}p{1cm}}
\hline
Psychological dimensions  & Symptom codes  & All   & Not being bullied   & Being bullied   & Chi-square value   & P-Value  \\
\midrule
\multirow{3}{*}{SDQ-S}  & 0   & 70848(74.2\%)  & 24288(89.6\%) & 46560(68.1\%) & \multirow{3}{*}{13.9}  & \multirow{3}{*}{***} \\
 & 1  & 13754(14.4\%)  & 1774(6.5\%)   & 11980(17.5\%) &  &  \\
& 2   & 10853(11.4\%)  & 1051(3.9\%)   & 9802(14.3\%)  &  &   \\ \hline
\multirow{2}{*}{PQ-16}     & 0 & 79744(83.5\%)  & 25892(95.5\%) & 53825(78.8\%) & \multirow{2}{*}{17.9}  & \multirow{2}{*}{***} \\
                                  & 1 & 15711(16.5\%)  & 1221(4.5\%)   & 14490(21.2\%) &                        &                                   \\ \hline
\multirow{4}{*}{GAD-7}        & 0      & 48749(51.1\%)  & 21491(79.3\%) & 27258(39.9\%) & \multirow{4}{*}{33.4}  & \multirow{4}{*}{***} \\
                                  & 1       & 34374(36.0\%)  & 4841(17.9\%)  & 29533(43.2\%  &                        &                                   \\
                                  & 2       & 9101(9.5\%)    & 612(2.3\%)    & 8489(12.4\%)  &                        &                                   \\
                                  & 3      & 3231(3.4\%)    & 169(0.6\%)    & 3067(4.5\%)   &                        &                                   \\ \hline
\multirow{3}{*}{CRIES-13}   & 0         & 73861(77.4\%)  & 24857(91.7\%) & 49004(71.7\%) & \multirow{2}{*}{16.3} & \multirow{3}{*}{***} \\
                                  & 1         & 16932(17.7\%)  & 1868(6.9\%)   & 15064(22.0\%) &                        &                                   \\
                                  & 2         & 4661(4.9\%)    & 387(1.4\%)    & 4274(6.3\%)   &                        &                                   \\ \hline
\multirow{4}{*}{PSQI}    & 0     & 64312(67.4\%)  & 23798(87.8\%) & 40514(59.3\%) & \multirow{4}{*}{21.0}    & \multirow{4}{*}{***} \\
                                  & 1     & 25326(26.5\%)  & 2884(10.6\%)  & 22442(32.8\%) &                        &                                   \\
                                  & 2   & 5366(5.6\%)    & 395(1.5\%)    & 4971(7.3\%)   &                        &                                   \\
                                  & 3     & 451(0.5\%)     & 36(0.1\%)     & 415(0.6\%)    &                        &                                   \\ \hline
\multirow{2}{*}{IGDS9-SF} & 0  & 88093(92.3\%)  & 26468(97.6\%) & 61625(90.2\%) & \multirow{2}{*}{4.7}   & \multirow{2}{*}{**}  \\
                                  & 1    & 7362(7.7\%)    & 654(2.4\%)    & 6717(9.8\%)   &                        &                                   \\ \hline
\multirow{5}{*}{PHQ-9} & 0 & 56108(58.8\%)  & 21649(79.8\%) & 34459(50.4\%) & \multirow{5}{*}{19.4}  & \multirow{5}{*}{***} \\
                                  & 1    & 23786(24.9\%)  & 4227(15.6\%)  & 19559(28.6\%) &                        &                                   \\
                                  & 2    & 9963(10.4\%)   & 877(3.2\%)    & 9086(13.3\%)  &                        &                                   \\
                                  & 3   & 3838(4.0\%)    & 269(1.0\%)    & 3569(5.2\%)   &                        &                                   \\
                                  & 4    & 1760(1.8\%)    & 91(0.3\%)     & 1669(2.4\%)   &                        &                                   \\ \hline
\multirow{4}{*}{WEMWBS}     & 0  & 12385(13.0\%) & 2671(9.9\%)   & 9714(14.2\%)  & \multirow{4}{*}{20.2}  & \multirow{4}{*}{***} \\
                                  & 1   & 14367(15.1\%)  & 1715(6.3\%)   & 12652(18.5\%) &                        &                                   \\
                                  & 2   & 48620(50.9\%)  & 11995(44.3\%) & 36625(53.6\%) &                        &                                   \\
                                  & 3   & 20083(21.0\%)  & 10732(39.6\%) & 9351(13.7\%)  &                        &   \\
\hline
\end{tabularx}
\end{table}

\subsection{The Impact of Bullying Severity}
After establishing significant correlations between bullying and mental health, we further analyzed correlations between the bullying severity and the presence of psychological problems. As shown in Table \ref{table2}, students who experienced mild bullying and students who experienced severe bullying significantly differed in terms of psychological conditions. The chi-square tests revealed that the severity of bullying is significantly correlated with seven psychological dimensions, except for the mental well-being, demonstrating that a higher severity of bullying experiences is associated with poorer psychological conditions.
\begin{table}[H]
\renewcommand{\arraystretch}{1.5}
\caption{The impact of the severity of bullying on psychological conditions.\label{table2}}
\newcolumntype{C}{>{\centering\arraybackslash}X}
\begin{tabularx}{\textwidth}{p{2cm}p{1.5cm}p{2cm}p{2cm}p{2cm}p{2cm}p{1cm}}
\toprule
Psychological dimensions & Symptom codes & Bullied & Mildly bullied & Severely bullied & Chi-square value  & P-Value   \\
\midrule
\multirow{3}{*}{SDQ-S}  & 0 & 46560(68.1\%) & 41401(73.4\%) & 5159(43.1\%) & \multirow{3}{*}{20.8} & \multirow{3}{*}{***} \\
                                  & 1         & 11980(17.5\%) & 9045(16.0\%)  & 2935(24.5\%) &                       &                                   \\
                                  & 2         & 9802(14.3\%)  & 5926(10.5\%)  & 3876(32.4\%) &          &    \\ \hline
\multirow{2}{*}{PQ-16} & 0 & 53825(78.8\%) & 47300(83.9\%) & 6552(54.7\%) & \multirow{2}{*}{20}   & \multirow{2}{*}{***} \\
                                  & 1  & 14490(21.2\%) & 9072(16.1\%)  & 5418(45.3\%) &     &    \\ \hline
\multirow{4}{*}{GAD-7}  & 0  & 27258(39.9\%) & 25210(44.7\%) & 2048(17.1\%) & \multirow{4}{*}{25.1} & \multirow{4}{*}{***} \\
  & 1       & 29533(43.2\%) & 24053(42.7\%) & 5479(45.8\%) &                       &                                   \\
                                  & 2      & 8489(12.4\%)  & 5596(9.9\%)   & 2893(24.2\%) &                       &                                   \\
                                  & 3   & 3067(4.5\%)   & 1512(2.7\%)   & 1550(12.9\%) &                       &  \\ \hline
\multirow{3}{*}{CRIES-13}   & 0         & 49004(71.7\%) & 43419(77.0\%) & 5585(46.7\%) & \multirow{3}{*}{21.1} & \multirow{3}{*}{***} \\
                                  & 1         & 15064(22.0\%) & 10767(19.1\%) & 4297(35.9\%) &       &   \\
                                  & 2         & 4274(6.3\%)   & 2186(3.9\%)   & 2088(17.4\%) &         &   \\ \hline
\multirow{4}{*}{PSQI}    & 0     & 40514(59.3\%) & 35957(63.8\%) & 4557(38.1\%) & \multirow{4}{*}{16}   & \multirow{4}{*}{**} \\
                                  & 1     & 22442(32.8\%) & 17299(30.7\%) & 5145(43.0\%) &                       &     \\
                                  & 2   & 4971(7.3\%)   & 2957(5.2\%)   & 2014(16.8\%) &                       &     \\
                                  & 3    & 415(0.6\%)    & 159(0.3\%)    & 256(2.1\%)   &                       &   \\ \hline
\multirow{2}{*}{IGDS9-SF} & 0    & 61625(90.2\%) & 52370(92.9\%) & 9255(77.3\%) & \multirow{2}{*}{9.5}  & \multirow{2}{*}{**} \\
                                  & 1     & 6717(9.8\%)   & 4002(7.1\%)   & 2715(22.7\%) &                       &  \\ \hline
\multirow{5}{*}{PHQ-9}          & 0        & 34459(50.4\%) & 29364(52.1\%) & 5095(42.6\%) & \multirow{5}{*}{10.2} & \multirow{5}{*}{**}  \\
                                  & 1      & 19559(28.6\%) & 17020(30.2\%) & 2539(21.2\%) &                       &    \\
                                  & 2      & 9086(13.3\%)  & 6856(12.2\%)  & 2230(18.6\%) &                       &  \\
                                  & 3     & 3569(5.2\%)   & 2321(4.1\%)   & 1248(10.4\%) &                       &   \\
                                  & 4    & 1669(2.4\%)   & 811(1.4\%)    & 858(7.2\%)   &                       &  \\ \hline
\multirow{4}{*}{WEMWBS}     & 0  & 9714(14.2\%)  & 6761(12.0\%)  & 2953(24.7\%) & \multirow{4}{*}{7.2}  &\multirow{4}{*}{*}   \\
                                  & 1 & 12652(18.5\%) & 9949(17.6\%)  & 2703(22.6\%) &                       &   \\
                                  & 2  & 36625(53.6\%) & 31478(55.8\%) & 5147(43.0\%) &                       &  \\
                                  & 3   & 9351(13.7\%)  & 8184(14.5\%)  & 1167(9.7\%)  &                       &  \\
\bottomrule
\end{tabularx}
\end{table}
\unskip
\par
To summarize the results from Tables \ref{table1} and \ref{table2}, students who experienced school bullying had generally poorer psychological conditions and a significantly higher risk of developing mental illnesses compared to students who did not experience it. For example, the proportion of students at risk of prodromal psychosis was 4.5\% among those who never experienced school bullying, while it was 21.2\% among those who experienced it, with 16.1\% and 45.3\% for those who experienced mild and severe bullying, respectively. All dimensions of psychological problems, with the exception of mental well-being, showed a consistent pattern of a higher risk of developing mental illness in those who experienced more severe bullying. 
\subsection{The Impact of Different Types of Bullying}
Among all participants, 30,720, 57,104, 43,028, and 54,729 subjects reported experiencing physical victimization, verbal victimization, social manipulation, and attacks on property, respectively, accounting for 44.9\%, 83.6\%, 73.7\%, and 80\% of all victims. Some participants reported experiencing multiple types of bullying. However, as shown in Table \ref{table_type}, there was no significant difference in the psychological conditions of victims in terms of the type of bullying. The chi-square tests confirmed no significant correlation between the type of school bullying and psychological conditions. In a word, the occurrence of psychological problems among victims is related to the extent of their exposure to bullying but not the type of bullying.

\begin{table}[H]\footnotesize
\renewcommand{\arraystretch}{1.5}
\setlength{\tabcolsep}{0.5mm}
\caption{The impact of different types of bullying on psychological problems.\label{table_type}}
\begin{tabularx}{\textwidth}{p{1.8cm}p{1.2cm}p{1.8cm}p{2cm}p{1.8cm}p{1.8cm}p{1.8cm}p{1.8cm}p{0.5cm}p{1cm}}
\toprule
Psychological dimensions & Symptom codes & All  & Physical victimization & Verbal victimization & Social manipulation & Attacks on property & Chi-square value & P-Value  \\ \hline
\multirow{3}{*}{SDQ-S}        & 0         & 70848(74.2\%)  & 17887(58.2\%) & 37145(65.0\%) & 25930(60.3\%) & 35923(65.6\%) & \multirow{3}{*}{1.72} & \multirow{3}{*}{\textgreater{}0.99} \\
                                  & 1         & 13754(14.4\%)   & 6567(21.4\%)    & 10714(18.8\%) & 8918(20.7\%)  & 10072(18.4\%) &                       &                                     \\
                                  & 2         & 10853(11.4\%) & 6266(20.4\%)    & 9245(16.2\%)  & 8180(19.0\%)  & 8734(16.0\%)  &     &   \\ \hline
\multirow{2}{*}{PQ-16}     & 0 & 79744(83.5\%) & 21399(69.7\%)   & 43567(76.3\%) & 30094(71.9\%) & 41736(76.3\%) & \multirow{2}{*}{1.67} & \multirow{2}{*}{\textgreater{}0.99} \\
                                  & 1  & 15711(16.5\%)  & 9321(30.3\%)    & 13537(23.7\%) & 12094(28.1\%) & 12993(23.7\%) & &   \\ \hline
\multirow{4}{*}{GAD-7}        & 0  & 48749(51.1\%) & 9515(31.0\%)    & 20713(36.3\%) & 13065(30.4\%) & 19685(36.0\%) & \multirow{4}{*}{1.54} & \multirow{4}{*}{\textgreater{}0.99} \\
                                  & 1    & 34374(36.0\%)  & 14214(46.3\%)   & 25630(44.9\%) & 20322(47.2\%) & 24691(45.1\%) &                       &                                   \\
                                  & 2      & 9101(9.5\%)   & 5006(16.3\%)    & 7859(13.8\%)  & 6972(16.2\%)  & 7532(17.2\%)  &                       &                                     \\
                                  & 3     & 3231(3.4\%)  & 1985(6.5\%)     & 2902(5.1\%)   & 2669(6.2\%)   & 2821(5.2\%)   &   &   \\ \hline
\multirow{3}{*}{CRIES-13}   & 0         & 73861(77.4\%) & 19019(61.9\%)   & 39285(68.8\%) & 27645(64.2\%) & 37705(68.9\%) & \multirow{3}{*}{1.96} & \multirow{3}{*}{\textgreater{}0.99} \\
                                  & 1         & 16932(17.7\%) & 8664(28.2\%)    & 13735(24.1\%) & 11631(27.0\%) & 13129(24.0\%) &                       &                                     \\
                                  & 2         & 4661(4.9\%) & 3037(9.9\%)     & 4084(7.2\%)   & 3752(8.7\%)   & 3895(7.1\%)   &      &    \\ \hline
\multirow{4}{*}{PSQI}    & 0    & 64312(67.4\%) & 15795(51.4\%)   & 32281(56.5\%) & 22087(51.3\%) & 30625(56.0\%) & \multirow{4}{*}{1.16} & \multirow{4}{*}{\textgreater{}0.99} \\
                                  & 1   & 25326(26.5\%) & 11439(37.2\%)   & 19785(34.6\%) & 16347(38.0\%) & 19204(35.1\%) &                       &                                     \\
                                  & 2     & 5366(5.6\%)   & 3168(10.3\%)    & 4637(8.1\%)   & 4211(9.8\%)   & 4509(8.2\%)   &                       &                                     \\
                                  & 3   & 451(0.5\%)   & 318(1.0\%)      & 401(0.7\%)    & 383(0.9\%)    & 391(0.7\%)    &    &    \\ \hline
\multirow{2}{*}{IGDS9-SF} & 0    & 88093(92.3\%)  & 26148(85.1\%)   & 50765(88.9\%) & 37451(87.0\%) & 48696(89.0\%) & \multirow{2}{*}{0.93} & \multirow{2}{*}{\textgreater{}0.8}  \\
                                  & 1   & 7362(7.7\%)     & 4572(14.9\%)    & 6339(11.1\%)  & 5577(13.0\%)  & 6023(11.0\%)  &      &     \\ \hline
\multirow{5}{*}{PHQ-9}          & 0        & 56108(58.8\%)  & 15363(50.0\%)   & 27628(48.4\%) & 19527(45.4\%) & 26135(47.8\%) & \multirow{5}{*}{1.04} & \multirow{5}{*}{1}                  \\
                                  & 1      & 23786(24.9\%)    & 7709(25.1\%)    & 16399(28.7\%) & 12195(28.3\%) & 16003(29.2\%) &                       &                                     \\
                                  & 2   & 9963(10.4\%)   & 4486(14.6\%)    & 8178(14.3\%)  & 6921(16.1\%)  & 7867(14.4\%)  &                       &                                     \\
                                  & 3    & 3838(4.0\%)   & 2062(6.7\%)     & 3313(5.8\%)   & 2914(6.8\%)   & 3179(5.8\%)   &                       &                                     \\
                                  & 4     & 1760(1.8\%)    & 1100(3.6\%)     & 1586(2.8\%)   & 1471(3.4\%)   & 1545(2.8\%)   &    &  \\ \hline
\multirow{4}{*}{WEMWBS}     & 0  & 12385((13.0\%)  & 5758(18.7\%)    & 8804(15.4\%)  & 7449(17.3\%)  & 8160(14.9\%)  & \multirow{4}{*}{1.15} & \multirow{4}{*}{\textgreater{}0.99} \\
                                  & 1  & 14367(15.1\%)    & 6525(21.2\%)    & 11187(19.6\%) & 9192(21.4\%)  & 10658(19.5\%) &                       &                                     \\
                                  & 2  & 48620(50.9\%)   & 15084(49.1\%)   & 30026(52.6\%) & 21881(50.9\%) & 29173(53.3\%) &                       &                                     \\
                                  & 3  & 20083(21.0\%) & 3353(10.9\%)    & 7087(12.4\%)  & 4508(10.5\%)  & 6738(12.3\%)  & &   \\
			\bottomrule
		\end{tabularx}
\end{table}

\subsection{Differences in School Bullying at Different Grade Levels}

Based on the grade levels (primary, junior high, or senior high school) of the students, we analyzed the differences in the proportion of students who experienced school bullying, as well as the types and degrees of bullying. As shown in Table 6, the proportions of students in primary school, junior high school, and high school who experienced school bullying were 71.6\%, 67.4\%, and 71.8\%, respectively. As confirmed by the chi-square tests, there were no significant differences between the three groups.
\par
In primary school, 16,470 (60.7\%) students experienced physical victimization. In contrast, the proportions of such students in junior high school and high school were significantly lower, at 32.3\% and 24.4\%, respectively. A significantly higher proportion of primary school students experienced severe bullying than junior high school and senior high school students, and a significantly higher proportion of junior high school students experienced severe bullying than senior high school students. The chi-square tests validated the statistical significance of the above two correlations. Based on these findings, we recommend paying particular attention to school bullying incidents in primary schools, especially those involving physical victimization.

\begin{table}[H]\footnotesize
\renewcommand{\arraystretch}{1.5}
\caption{The difference in bullying prevalence among different grade levels.\label{table_date}}
  \begin{tabularx}{\textwidth}{p{1.8cm}p{1.8cm}p{1.5cm}p{1.5cm}p{1.8cm}p{1.8cm}p{1.5cm}p{1cm}}
			\toprule
  & Category & Members & Primary School & Junior High School & Senior High School & Chi-Square Value & P-value \\
			\midrule
\multirow{2}{*}{\parbox{2cm}{Whether being bulling}}& Being bullied  & 68342(71.6\%) & 18271(67.4\%) & 31964(74.1\%) & 18107(71.8\%) & \multirow{2}{*}{1.1} & \multirow{2}{*}{0.56} \\
 & Not being bullied & 27113(28.4\%) & 8857(32.6\%)  & 11160(25.9\%) & 7096(28.2\%)  &  &  \\ \hline
 \multirow{4}{*}{\parbox{2cm}{Types of bullying}}& Physical victimization  & 30720(32.2\%) & 16470(60.7\%) & 13923(32.3\%) & 6139(24.4\%)  & \multirow{4}{*}{16}  & \multirow{4}{*}{**}   \\
 & Verbal victimization  & 57104(59.8\%) & 15530(57.2\%) & 27194(63.1\%) & 14380(57.1\%) &     &  \\
& Social manipulation & 43028(45.1\%) & 12327(45.4\%) & 20080(46.6\%) & 10621(42.1\%) &    & \\
& Attacks on property  & 54729(57.3\%) & 14495(53.4\%) & 25663(59.5\%) & 14571(57.8\%) &   & \\ \hline
\multirow{2}{*}{\parbox{2cm}{Severity of bullying }}  & Mild bullying & 56372(82.5\%) & 14160(77.5\%) & 26526(83\%)   & 15956(88.1\%) & \multirow{2}{*}{3.9} & \multirow{2}{*}{**}   \\
& Severe bullying & 11970(17.5\%) & 4111(22.5\%)  & 5708(17\%)    & 2151(11.9\%)  &      &  \\
			\bottomrule
		\end{tabularx}
\end{table}

\subsection{Relationship Between School Bullying and the Probability of Psychological Illnesses}
Having established the significant correlations between school bullying experience and psychological problems, we sought to demonstrate the impact of school bullying experience on the risk of developing psychological illnesses in a more intuitive way. To achieve this, we divided the participants into two groups: those with and those without the psychological illness for each dimension, according to Table \ref{table_class}. Next, we performed a logistic regression on each dimension. Table \ref{table3} presents the results, with the odds ratio indicating the likelihood of having the corresponding psychological illness for the group that experienced school bullying compared to the group that did not. For example, as shown in Table \ref{table3}, the odds of developing emotional and behavioral disorders for students who experienced mild and severe school bullying were 3.10 times and 11.35 times higher, respectively, than for students who did not experience bullying. These results indicate that experiencing school bullying significantly increases the probability of developing psychological illnesses, and this probability increases with the severity of bullying.

\begin{table}[H] 
\renewcommand{\arraystretch}{1.5}
\caption{Regression results of the 8 dimensions of psychological problems.\label{table3}}
\newcolumntype{C}{>{\centering\arraybackslash}X}
\begin{tabularx}{\textwidth}{p{2cm}p{2.5cm}p{2cm}p{1.5cm}p{2.5cm}C}
\toprule
Psychological dimensions & Severity & Reference value & Odds Ratio (OR) & 95\% Confidence Interval & P-Value    \\ 
\midrule
\multirow{2}{*}{SDQ-S}   & Mild bullying & \multirow{2}{*}{No bullying} & 3.10& 2.97-3.24     & ***      \\
                         & Severe bullying &                            & 11.35 & 10.76-11.97   & ***   \\ \midrule
\multirow{2}{*}{PQ-16}     & Mild bullying             & \multirow{2}{*}{No bullying} & 4.06                 & 3.82-4.32            & ***      \\
                                  & Severe bullying             &                         & 17.53                & 16.38-18.76          & ***      \\ \midrule
\multirow{2}{*}{GAD-7}        & Mild bullying             & \multirow{2}{*}{No bullying} & 4.72                 & 4.56-4.88            & ***      \\ 
                                  & Severe bullying             &                         & 18.52                & 17.51-19.58          & ***      \\ \midrule
\multirow{2}{*}{CRIES-13}   & Mild bullying             & \multirow{2}{*}{No bullying} & 3.28                 & 3.13-3.44            & ***      \\
                                  & Severe bullying             &                         & 12.59                & 11.90-13.32          & ***      \\ \midrule
\multirow{2}{*}{PSQI}    & Mild bullying             & \multirow{2}{*}{No bullying} & 4.07                 & 3.91-4.24            & ***      \\
                                  & Severe bullying             &                         & 11.67                & 11.08-12.29          & ***      \\ \midrule
\multirow{2}{*}{IGDS9-SF} & Mild bullying             & \multirow{2}{*}{No bullying} & 3.13                 & 2.88-3.41            & ***      \\
                                  & Severe bullying             &                         & 12.03                & 11.01-13.15          & ***      \\ \midrule
\multirow{2}{*}{PHQ-9}          & Mild bullying             & \multirow{2}{*}{No bullying} & 3.64  & 3.52-3.77            & ***      \\
                                  & Severe bullying             &                         & 5.34                 & 5.10-5.60            & ***      \\ \midrule
\multirow{2}{*}{WEMWBS}     & Mild bullying             & \multirow{2}{*}{No bullying} & 2.18                 & 2.10-2.26            & ***      \\
                                  & Severe bullying             &                         & 4.64                 & 4.42-4.87            & ***      \\

\bottomrule
\end{tabularx}
\end{table}
\unskip

\section{Conclusion and Discussion}
In this study, we investigated the status of school bullying among primary and secondary school students in Zigong City. The results showed that of the 95,545 surveyed students, 71.6\% (\textit{68,342}) experienced school bullying of varying degrees, and among students who were bullied, 17.5\% (\textit{11,970}) experienced severe bullying. This indicates that school bullying is a prevalent issue among primary and secondary school students, with a higher incidence among the former. Additionally, the severity is more pronounced among primary school students than their secondary school counterparts. There are notable differences in the prevalence of school bullying across different countries. In a previous study, the percentage of students involved in school bullying across different countries ranged from 6.3\% to 45.2\% \cite{6}. In comparison, our investigation reported much higher percentage. This difference may be attributed to the relatively higher levels of social development and implementation of preventive laws and regulations against school bullying in the European countries. Meanwhile, China, as a developing country, and Zigong, as an underdeveloped city in china, need more efforts to address this issue. 
\par
The chi-square tests and logistic regression showed that the occurrence and severity of school bullying are significantly correlated with all eight psychological dimensions. Compared with students who never experienced school bullying, students who experienced school bullying were more likely to have psychological problems, and the severity of bullying was positively correlated with a higher likelihood of psychological problems. These findings are consistent with previous research, which suggests that victims of school bullying may struggle to solve life problems and often have negative attitudes and poor interpersonal relationships \cite{22}. Other studies indicate that the social pressure experienced by victims may result in a strong sense of threat, which may lead to psychological problems such as depression, anxiety, fear of attending school, and feelings of insecurity and dissatisfaction in school \cite{23}. Additionally, the vigilance and stress response of teenagers during puberty may increase significantly, potentially increasing the risk of psychological problems in the conflicting environment \cite{24}. 
\par
Studying the relationship between school bullying and psychological conditions has important social value and practical significance. It provides a better understanding of the origin of school bullying and how school bullying impacts students' mental health, thereby offering a scientific basis for preventing and reducing school bullying. It also helps in treating bullying-related psychological problems.
\par
School bullying is a serious social problem that considerably affects the physical and mental health of victims, leading to emotional and behavioral problems, anxiety, depression, PTSD, and even extreme behaviors such as suicide. To reduce the occurrence of school bullying, we propose the following suggestions. First, school administrators, teachers, and parents should dearly understand what bullying is and its impacts on the physical and mental health. They should strengthen education on preventing bullying, closely monitor students, and pay attention to students’ daily lives. Second, both society and schools should take bullying seriously, promptly stop bullying incidents, and provide necessary help to students with psychological problems. Thirdly, after a bullying incident occurs, school administrators and parents should immediately pay attention to the safety of the victims, dig out the reason leading to the bullying incident, figure out the detailed process and property of the bullying incident, help the victims overcome their psychological trauma, improve their self-protection abilities, and prevent further occurrences. Fourth, educational managers and school administrators should prioritize creating a safer and more inclusive campus. Preventing school bullying should be a core part of developing a safe school. Technological means, routine surveys, and trained inquiry methods should be used to promptly identify and address potential school bullying incidents. 
\par
The present study has several strengths worth mentioning. First, the sample size was large, with 95,545 students surveyed, providing significant and stable statistical results. Second, the study was conducted in a developing country, making the findings more relevant to other developing countries. Third, the study differentiated between the four types  and the severity of school bullying, and thus drew detailed conclusions. Finally, the study utilized multiple scales and questionnaires to measure various psychological problems, allowing for a comprehensive analysis of the quantitative relationships between school bullying and various psychological illnesses.
\par
This study also has certain limitations. Firstly, the cross-sectional design precludes causal inferences about the relationship between school bullying and psychological problems and only suggests correlations. The focus of our future research should be on establishing causal relationships between school bullying and psychological problems through longitudinal studies. Secondly, the data were self-reported by the students, which may have subjective biases, potentially overestimating or underestimating the correlations. Future research should employ more objective measures to collect data on students’ psychological issues and bullying behaviors. Thirdly, our research focused solely on a specific city, and while it may have some representativeness, the comparative analysis of other Chinese cities during the same period is insufficient. Finally, it is worth noting that the occurrence of school bullying and psychological illnesses may not be evenly distributed among the population, which may have led to skewed probability estimates. Moreover, the COVID-19 pandemic may have largely impacted the findings. Therefore, future studies should consider more in-depth comparative analysis, using data from post-pandemic investigations, to provide a more accurate understanding of the relationship between school bullying and psychological problems among school students.

\end{document}